\documentclass[%
 reprint,
nofootinbib,
 amsmath,amssymb,
 aps,
]{revtex4-1}

\usepackage{graphicx}
\usepackage{dcolumn}
\usepackage{bm}

\usepackage{ulem}  

\begin{document}




\title{Shadowing in inelastic nucleon-nucleon cross section?}



\author{Kari J. Eskola}%
\email{kari.eskola@jyu.fi}
\author{Ilkka Helenius}%
\email{ilkka.m.helenius@jyu.fi}
\author{Mikko Kuha}%
\email{mikko.a.kuha@student.jyu.fi}
\author{Hannu Paukkunen}%
\email{hannu.t.paukkunen@jyu.fi}
\affiliation{%
University of Jyvaskyla, Department of Physics, P.O. Box 35, FI-40014 University of Jyvaskyla, Finland}
\affiliation{%
Helsinki Institute of Physics, P.O. Box 64, FI-00014 University of Helsinki, Finland}%

\date{\today}

\begin{abstract}

Experimental results of inclusive hard-process cross sections in heavy-ion collisions conventionally lean on a normalization computed from Glauber models where the inelastic nucleon-nucleon cross section $\sigma_{\rm nn}^{\rm inel}$ -- a crucial input parameter -- is simply taken from proton-proton measurements. In this letter, using the computed electro-weak boson production cross sections in lead-lead collisions as a benchmark, we determine $\sigma_{\rm nn}^{\rm inel}$ from the recent ATLAS data. We find a significantly suppressed $\sigma_{\rm nn}^{\rm inel}$ relative to what is usually assumed, show the consequences for the centrality dependence of the cross sections, and address the phenomenon in an eikonal minijet model with nuclear shadowing.

\end{abstract}

\maketitle



\section{Introduction}
\label{sec:intro}

In a high-energy nucleus-nucleus collision the produced particle multiplicity correlates strongly with the collision geometry: the more central the collision, typically the higher the multiplicity. Experimentally, the centrality classification is obtained by organizing the events according to their multiplicity (or transverse energy) into bins of equal fraction, say 10\%, of all events. Conventionally, 0--10\% (90--100\%) centrality refers to the events of highest (lowest) multiplicities, and 0--100\% to all events, minimum bias.

Inclusive hard processes in turn are rarer processes of a large momentum scale whose cross sections in nucleus-nucleus collisions are traditionally obtained by converting the measured per-event yields $N_h^c/{N_{\mathrm{evt}}^c}$ in a centrality class $c$ into hard nucleon-nucleon cross sections $\sigma_h^c$ through
\begin{equation}
\sigma_h^c = 
\frac{\sigma_{\mathrm{nn}}^{\mathrm{inel}}}{\langle N_{\rm bin} \rangle_c}
\frac{N_h^c}{N_{\mathrm{evt}}^c} \,,
\label{eq:crossexp}
\end{equation}
where $\langle N_{\rm bin} \rangle_c$ is the mean number of independent inelastically interacting nucleon-nucleon pairs, binary collisions, in the centrality class $c$, and $\sigma_{\mathrm{nn}}^{\mathrm{inel}}$ is the inelastic nucleon-nucleon cross section. The model-dependent quantity $\langle N_{\rm bin} \rangle_c$ here is obtained from the Monte-Carlo (MC) Glauber model \cite{Miller:2007ri}. The nuclear modification ratio $R_{AA}^{h,c}$ for the hard process, in the centrality class $c$, is then obtained by dividing $\sigma_h^c$ by the corresponding minimum-bias cross-section in proton-proton collisions.

This method of constructing hard cross sections is a routine procedure in the heavy-ion measurements at RHIC and at the LHC and it has been used e.g. to construct nuclear modification ratios for jets \cite{Abelev:2013kqa, Khachatryan:2016jfl, Aad:2012vca, Aaboud:2018twu} and charged particles \cite{Sirunyan:2018eqi, Khachatryan:2016odn, Aad:2015wga, Acharya:2018eaq, Acharya:2018qsh, Adare:2015bua, Adler:2003au} which, in turn, are widely used in theoretical studies of jet quenching \cite{Bjorken:1982tu, Qin:2015srf, dEnterria:2009xfs} and partonic energy loss \cite{Arleo:2017ntr, Andres:2016iys, Chien:2015vja, Zapp:2012ak}. In the same way, Eq.~(\ref{eq:crossexp}) forms the basis for measuring centrality-dependent cross sections of direct photon \cite{Aad:2015lcb, Chatrchyan:2012vq, Afanasiev:2012dg} and electro-weak (EW) boson \cite{Aad:2014bha, Aad:2012ew, Chatrchyan:2012nt, Chatrchyan:2011ua} production, which can be used to study e.g. nuclear effects in parton distribution functions (PDFs) \cite{Paukkunen:2010qg, Helenius:2013bya}. 

The basic inputs of the Glauber model are the nuclear geometry and $\sigma_{\mathrm{nn}}^{\mathrm{inel}}$ \cite{Miller:2007ri}. In the MC Glauber model the positions of the nucleons are sampled event by event according to the nuclear density profile, usually the Woods-Saxon distribution \cite{Woods:1954zz}. The probability for an interaction between two nucleons depends on their mutual distance and $\sigma_{\mathrm{nn}}^{\mathrm{inel}}$. As a result, cross-section measurements through Eq.~(\ref{eq:crossexp}) depend on $\sigma_{\mathrm{nn}}^{\mathrm{inel}}$ in a non-trivial way. An established procedure is to take the value $\sigma_{\mathrm{nn}}^{\mathrm{inel}}$ and its energy dependence from proton-proton measurements. However, at high-enough energies the particle production becomes sensitive to QCD dynamics at small momentum fractions $x$ where some suppression is expected due to gluon shadowing \cite{Qiu:1986wh, Armesto:2006ph, Frankfurt:2011cs} or saturation phenomena \cite{Gribov:1984tu, McLerran:1993ni, Gelis:2010nm}. Such effects become more pronounced in heavy nuclei and towards lower scales so one could argue that in collisions involving heavy ions the value of $\sigma_{\mathrm{nn}}^{\mathrm{inel}}$ should also be reduced relative to what is measured in proton-proton collisions. Through Eq.~(\ref{eq:crossexp}), this would then change the obtained hard cross sections and nuclear modification ratios, and thereby affect all the subsequent analyses that take these measured cross sections as an input. In this way, the value of $\sigma_{\mathrm{nn}}^{\mathrm{inel}}$ could be critical and have far-reaching consequences e.g. for the precision studies of jet quenching and other related phenomena. Thus, an alternative benchmark for $\sigma_{\mathrm{nn}}^{\mathrm{inel}}$ is called for. 

As proposed in Ref.~\cite{Paukkunen:2010qg}, the Glauber model and its inputs could be tested by studying the production of well known ''standard candles'', such as EW bosons, in Pb+Pb collisions at the LHC, but so far this has been limited by the precision of the LHC Run-I measurements \cite{Aad:2012ew, Chatrchyan:2012nt, Aad:2014bha, Chatrchyan:2014csa}. Thanks to the increased luminosity and collision energy of Run II, the recent $W^{\pm}$- and $Z$-boson measurements by ATLAS \cite{Aad:2019sfe, Aad:2019lan} have pushed the precision to a few-percent level enabling now a more precise Glauber model calibration. In the present letter, we use these ATLAS data to study the possible nuclear suppression of $\sigma_{\mathrm{nn}}^{\mathrm{inel}}$ in Pb+Pb collisions. Since the ALICE measurement \cite{Acharya:2017wpf} is less precise and has no reference p+p data we leave it out from the analysis. The idea is to first nail down the EW-boson cross sections by using a next-to-next-to-leading order (NNLO) perturbative QCD (pQCD) with state-of-the-art PDFs for protons and nuclei. Using the theory prediction on the left-hand-side of Eq.~(\ref{eq:crossexp}), we can then determine $\sigma_{\mathrm{nn}}^{\mathrm{inel}}$ within the same MC Glauber implementation as in the experimental analyses. We find that the data favor a significant suppression in $\sigma_{\mathrm{nn}}^{\mathrm{inel}}$. We show that this is compatible with predictions from an eikonal minijet model with nuclear shadowing. We also demonstrate that the unexpected enhancement seen by ATLAS in the ratios $R_{\mathrm{PbPb}}^{W^\pm,Z}$ towards peripheral collisions disappears with the found smaller value of $\sigma_{\mathrm{nn}}^{\mathrm{inel}}$. 


\section{Nuclear suppression in $\sigma_{\mathrm{nn}}^{\mathrm{inel}}$}

The observables we exploit in this work to extract $\sigma_{\mathrm{nn}}^{\mathrm{inel}}$ are the rapidity-dependent nuclear modification ratios for $W^{\pm}$ and $Z$ boson production in different centrality classes. Experimentally these are defined as
\begin{equation}
R_{\mathrm{PbPb}}^{\mathrm{exp}}(y) = \frac{1}{\langle T_{AA}\rangle}
\frac{
\frac{1}{N_{\mathrm{evt}}}
\mathrm{d}N^{W^{\pm}, Z}_{\mathrm{PbPb}}/\mathrm{d} y}{\mathrm{d}\sigma^{W^{\pm}, Z}_{\mathrm{pp}}/\mathrm{d}y},
\label{eq:Rexp}
\end{equation}
where the per-event yield is normalized into nucleon-nucleon cross section by diving with the mean nuclear overlap $\langle T_{AA} \rangle = \langle N_{\rm bin} \rangle_c / \sigma_{\rm nn}^{\rm inel}$ obtained from a MC Glauber model calculation. For minimum-bias collisions the same quantity can be calculated directly as a ratio between the cross sections in Pb+Pb and p+p collisions, 
\begin{equation}
R_{\mathrm{PbPb}}^{\mathrm{theor}}(y) = \frac{1}{(208)^2} \frac{\mathrm{d}\sigma^{W^{\pm}, Z}_{\mathrm{PbPb}}/\mathrm{d} y}{\mathrm{d}\sigma^{W^{\pm}, Z}_{\mathrm{pp}}/\mathrm{d}y}.
\label{eq:Rtheor}
\end{equation}
We have calculated the cross sections in Eq.~(\ref{eq:Rtheor}) at NNLO with the \textsc{mcfm} code (version 8.3) \cite{Boughezal:2016wmq}. For the protons we use the recent NNPDF3.1 PDFs \cite{Ball:2017nwa} which provide an excellent agreement to ATLAS data for $W^{\pm}$ and $Z$ boson production in p+p collisions at $\sqrt{s}=5.02~\text{TeV}$ \cite{Aaboud:2018nic}. The nuclear modifications for the PDFs are obtained from the EPPS16 NLO analysis \cite{Eskola:2016oht} which includes Run-I data for $W^{\pm}$ and $Z$ production in p+Pb collisions at the LHC \cite{Khachatryan:2015hha, Khachatryan:2015pzs, Aad:2015gta} and provide an excellent description of the more recent Run-II data \cite{Sirunyan:2019dox}. The available NNLO nuclear PDFs \cite{Walt:2019slu, AbdulKhalek:2019mzd} do not include any constraints beyond deeply inelastic scattering, so the applied PDFs provide currently the most accurately constrained setup for the considered observables. The factorization and renormalization scales are fixed to the respective EW boson masses.

\begin{figure}
\center
\includegraphics[width=1.00\linewidth]{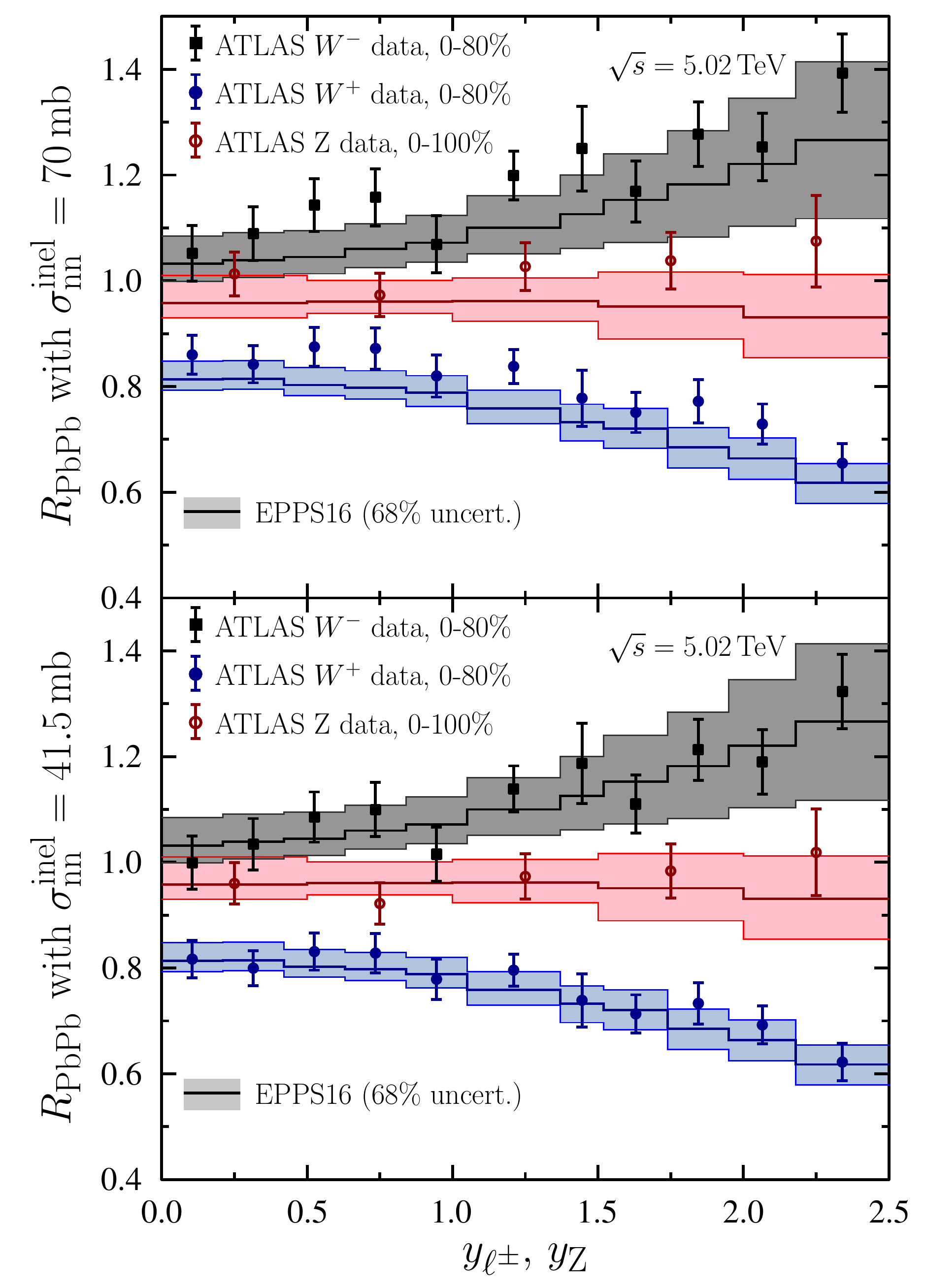}
\caption{Nuclear  modification ratios of $W^{\pm}$ and $Z$, computed from pQCD (solid lines with error bands)
and from ATLAS data \cite{Aad:2019sfe, Aad:2019lan} with $\sigma_{\mathrm{nn}}^{\mathrm{inel}} = 70 \, {\rm mb}$ (upper panel) and 41.5 mb (lower panel).}
\label{fig:R_PbPb}
\end{figure}

The ratios $R_{\mathrm{PbPb}}^{\mathrm{theor}}$ and $R_{\mathrm{PbPb}}^{\mathrm{exp}}$ are compared in the upper panel of Fig.~\ref{fig:R_PbPb}. For $W^{\pm}$, $R_{\mathrm{PbPb}}^{\mathrm{exp}}$ is formed by diving the normalized yield in Pb+Pb from Ref.~\cite{Aad:2019sfe} with the corresponding cross section in p+p from Ref.~\cite{Aaboud:2018nic} adding the uncertainties in quadrature. The plotted experimental uncertainties do not include the uncertainty in $\langle T_{AA} \rangle$. The theoretical uncertainties derive from the EPPS16 error sets and correspond to the 68\% confidence level. Note that the $W^{\pm}$ measurement is for 0--80\% centrality instead of full 0--100\%. However, for rare processes like the EW bosons the contribution from the 80--100\% region is negligible so the comparison with the minimum-bias calculations is justified. It is evident that with $\sigma_{\mathrm{nn}}^{\mathrm{inel}} = 70 \, {\rm mb}$ both the $W^{\pm}$ and the $Z$ data tend to lie above the calculated result, which we will interpret as an evidence of nuclear suppression in $\sigma_{\mathrm{nn}}^{\mathrm{inel}}$ as explained below.

\begin{figure}
\includegraphics[width=0.49\textwidth]{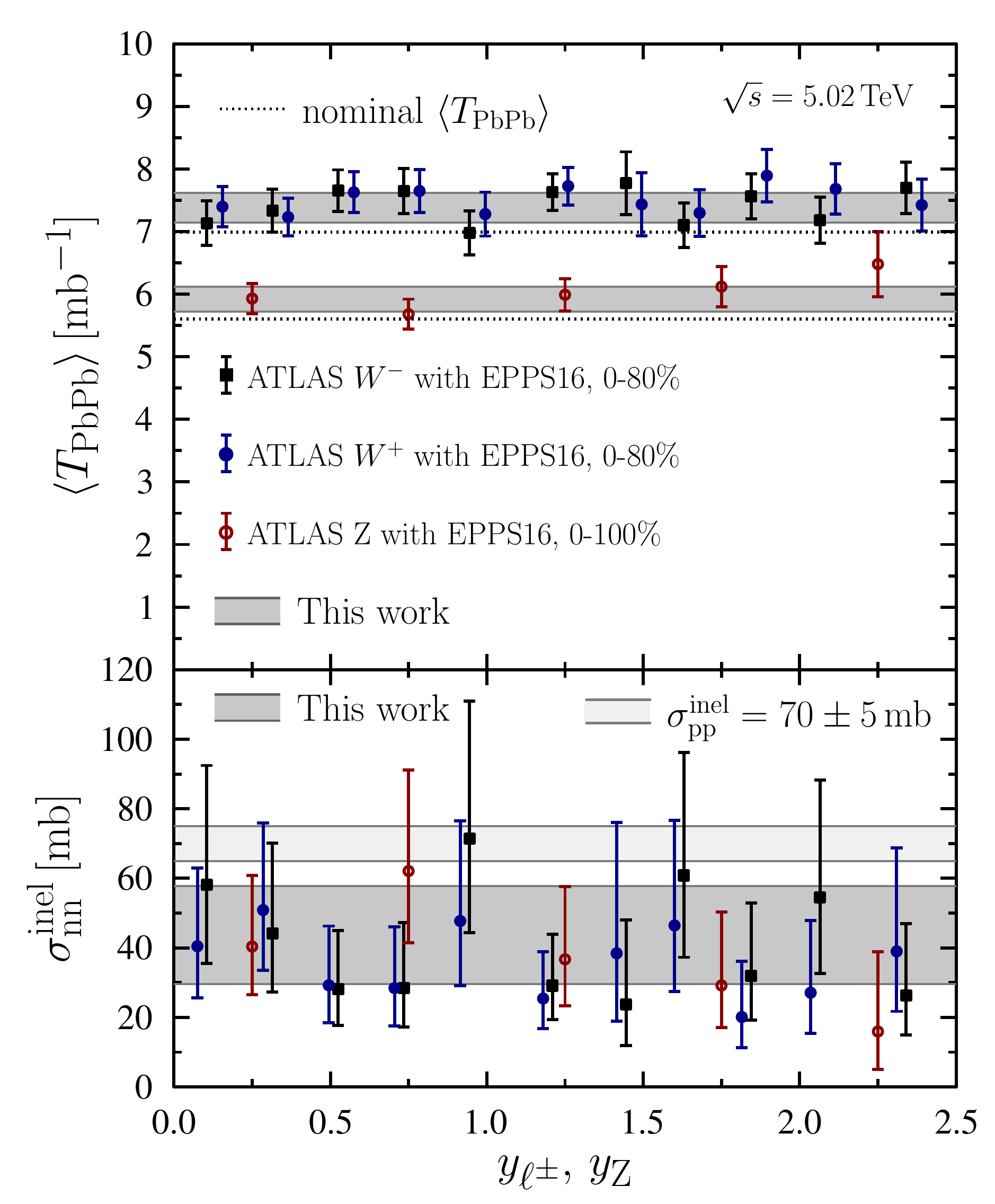}
\caption{Extracted values of the mean nuclear overlap functions (upper panel) and $\sigma_{\mathrm{nn}}^{\mathrm{inel}}$ (lower panel). The dark-gray bands show the values obtained by fitting $\sigma_{\mathrm{nn}}^{\mathrm{inel}}$ and the dashed lines and the light-gray band corresponds to the the nominal $\sigma_{\mathrm{pp}}^{\mathrm{inel}}$.}
\label{fig:TAA_sigma_fit}
\end{figure}

By equating Eqs.~(\ref{eq:Rexp}) and (\ref{eq:Rtheor}) we can convert each data point to $\langle T_{AA} \rangle$. The outcome is shown in the upper panel of Fig.~\ref{fig:TAA_sigma_fit}. The obtained values tend to be higher than the nominal $\langle T_{AA} \rangle = 5.605\,{\rm mb}^{-1}$ (0--100\%) and $\langle T_{AA} \rangle = 6.993\,{\rm mb}^{-1}$ (0--80\%) which assume $\sigma_{\mathrm{nn}}^{\mathrm{inel}} = 70 \, {\rm mb}$, see Table~\ref{tab:cent}. The fact that the preferred values of $\langle T_{AA} \rangle$ are independent of the rapidity strongly suggests that the original mismatch in $R_{\mathrm{PbPb}}$ is a normalization issue -- the nuclear PDFs predict the rapidity dependence correctly. 

Since each $\langle T_{AA} \rangle$ maps to $\sigma_{\mathrm{nn}}^{\mathrm{inel}}$ through MC Glauber, we can also directly convert $R_{\mathrm{PbPb}}^{\mathrm{exp}}$ to $\sigma_{\mathrm{nn}}^{\mathrm{inel}}$. Here, we have used \textsc{TGlauberMC} (version 2.4) \cite{Loizides:2014vua} which is the same MC Glauber implementation as in the considered ATLAS analyses. The centrality classification is done with a two-component model, including negative binomial fluctuations%
\footnote{While modifying $\sigma_{\mathrm{nn}}^{\mathrm{inel}}$, the parameters of the two-component model should be adjusted to maintain a good description of the measured multiplicity or transverse energy distribution. However, as the change in $\sigma_{\mathrm{nn}}^{\mathrm{inel}}$ can be accurately compensated purely by increasing the mean of the negative binomial distribution not affecting the resulting {$\langle T_{AA} \rangle $}, the presented results would remain unmodified.}%
, similar to the ALICE prescription \cite{Abelev:2013qoq} with parameters from Ref.~\cite{ALICE-PUBLIC-2018-011}. The obtained values of $\langle T_{AA} \rangle$ are in an excellent agreement with the ATLAS values in Refs.~\cite{Aad:2019sfe, Aad:2019lan} in all centrality classes when using the nominal, unsuppressed, value $\sigma_{\mathrm{nn}}^{\mathrm{inel}} = 70~\text{mb}$. The values of $\sigma_{\mathrm{nn}}^{\mathrm{inel}}$ extracted from each data point are shown in Fig.~\ref{fig:TAA_sigma_fit}. It is obvious that the data prefer a value of $\sigma_{\mathrm{nn}}^{\mathrm{inel}}$ which is less than the $\sigma_{\mathrm{pp}}^{\mathrm{inel}} = 70\,{\rm mb}$ obtained from p+p data. 

\begin{table}[htb!]
\caption{Mean nuclear overlap functions $\langle T_{AA} \rangle~[1/\text{mb}]$ for ATLAS centrality classes with nominal and fitted $\sigma_{\mathrm{nn}}^{\mathrm{inel}}$.}
\label{tab:cent}
\begin{ruledtabular}
\begin{tabular}{ccccc}
$\sigma_{\mathrm{nn}}^{\mathrm{inel}}$ & $70.0~\text{mb}$ & $57.7~\text{mb}$ & $41.5~\text{mb}$ & $29.5~\text{mb}$ \\
\hline
$   0 -  2\%$ & 28.26 & 28.39 & 28.55 & 28.69 \\ 
$   2 -  4\%$ & 25.51 & 25.67 & 25.91 & 26.10 \\ 
$   4 -  6\%$ & 23.09 & 23.28 & 23.55 & 23.80 \\ 
$   6 -  8\%$ & 20.94 & 21.14 & 21.45 & 21.73 \\ 
$   \phantom{1}8 - 10\%$ & 19.00 & 19.23 & 19.56 & 19.86 \\ 
$  10 - 15\%$ & 16.08 & 16.31 & 16.67 & 17.02 \\ 
$  15 - 20\%$ & 12.58 & 12.83 & 13.22 & 13.59 \\ 
$  20 - 25\%$ &  9.762 & 10.01 & 10.40 & 10.78 \\ 
$  25 - 30\%$ &  7.487 &  7.722 &  8.102 &  8.469 \\ 
$  30 - 40\%$ &  4.933 &  5.138 &  5.474 &  5.808 \\ 
$  40 - 50\%$ &  2.628 &  2.780 &  3.036 &  3.300 \\ 
$  50 - 60\%$ &  1.281 &  1.378 &  1.550 &  1.733 \\ 
$  60 - 80\%$ &  0.395 &  0.435 &  0.510 &  0.595 \\ 
$  \phantom{1}80 -100\%$ &  0.052 &  0.060 &  0.076 &  0.096 \\ 
\hline
$  \phantom{1}0 -80\%$ & 6.993  & 7.143 & 7.385 & 7.624 \\ 
$  \phantom{10}0 -100\%$ & 5.605  & 5.726 & 5.923 & 6.118 \\ 
\end{tabular}
\end{ruledtabular}
\end{table}

To quantify the optimal $\sigma_{\mathrm{nn}}^{\mathrm{inel}}$ we fit its value by requiring a match between $R_{\mathrm{PbPb}}^{\mathrm{exp}}$ and $R_{\mathrm{PbPb}}^{\mathrm{theor}}$ treating the EPPS16 uncertainties as Gaussian correlated errors. In practice we define a $\chi^2$ function by
\begin{eqnarray}
\chi^2 & = & \sum_{i} 
\left[
\frac{\mathcal{N}_i R_i^{\mathrm{exp}}- R_i^{\mathrm{theor}} + \sum_k f_k \beta_i^k}{\mathcal{N}_i\delta_i^{\mathrm{exp}} }
\right]^2 + T \sum_k f_k^2
\nonumber \\ 
\mathcal{N}_i &  = & {\langle T_{AA}^i(\sigma_{\mathrm{pp}}^{\mathrm{inel}})\rangle}/{\langle T_{AA}^i(\sigma_{\mathrm{nn}}^{\mathrm{inel}})\rangle}
\end{eqnarray}
where $i$ runs over the data points and $k=1,\ldots,20$ over the number error-set pairs in EPPS16. The factors $\mathcal{N}_i$ with $\sigma_{\mathrm{pp}}^{\mathrm{inel}} = 70~\text{mb}$ account for the shifted normalizations when $\sigma_{\mathrm{nn}}^{\mathrm{inel}}$ changes. Also the data uncertainties $\delta_i^{\mathrm{exp}}$ are scaled by this factor to avoid D'Agostini bias \cite{DAgostini:1993arp}. The tolerance $T=1.645^2$ in the penalty term takes into account scaling the 90\% confidence limit uncertainties of EPPS16 into 68\% and
$%
\beta_i^k \equiv \left[ R_i^{\mathrm{theor}}(S_k^+) - R_i^{\mathrm{theor}}(S_k^-)\right]/2,
$%
where $S_k^+$ and $S_k^-$ are the positive and negative variations, respectively, of EPPS16 error sets. The $\chi^2$ is minimized with respect to $\sigma_{\mathrm{nn}}^{\mathrm{inel}}$ and $f_k$ (1+20 parameters). We find 
$$\sigma_{\mathrm{nn}}^{\mathrm{inel}} = 41.5^{+16.2}_{-12.0}~\text{mb}\,,$$ where the uncertainties follow from the $\Delta \chi^2 = 1$ criterion. The resulting values for $\langle T_{AA} \rangle$ and $\sigma_{\mathrm{nn}}^{\mathrm{inel}}$ are compared to the data-extracted values in Fig.~\ref{fig:TAA_sigma_fit}, and the re-normalized data for $R_{\mathrm{PbPb}}$ are compared with theoretical predictions in the lower panel of Fig.~\ref{fig:R_PbPb}. It is worth stressing that different final states prefer a very similar, suppressed value of $\sigma_{\mathrm{nn}}^{\mathrm{inel}}$ and that a very good agreement in $R_{\mathrm{PbPb}}$ is found when normalizing with $\langle T_{AA} \rangle$ calculated using the suppressed cross section in the MC Glauber calculation.

\begin{figure*}
\includegraphics[width=0.49\textwidth]{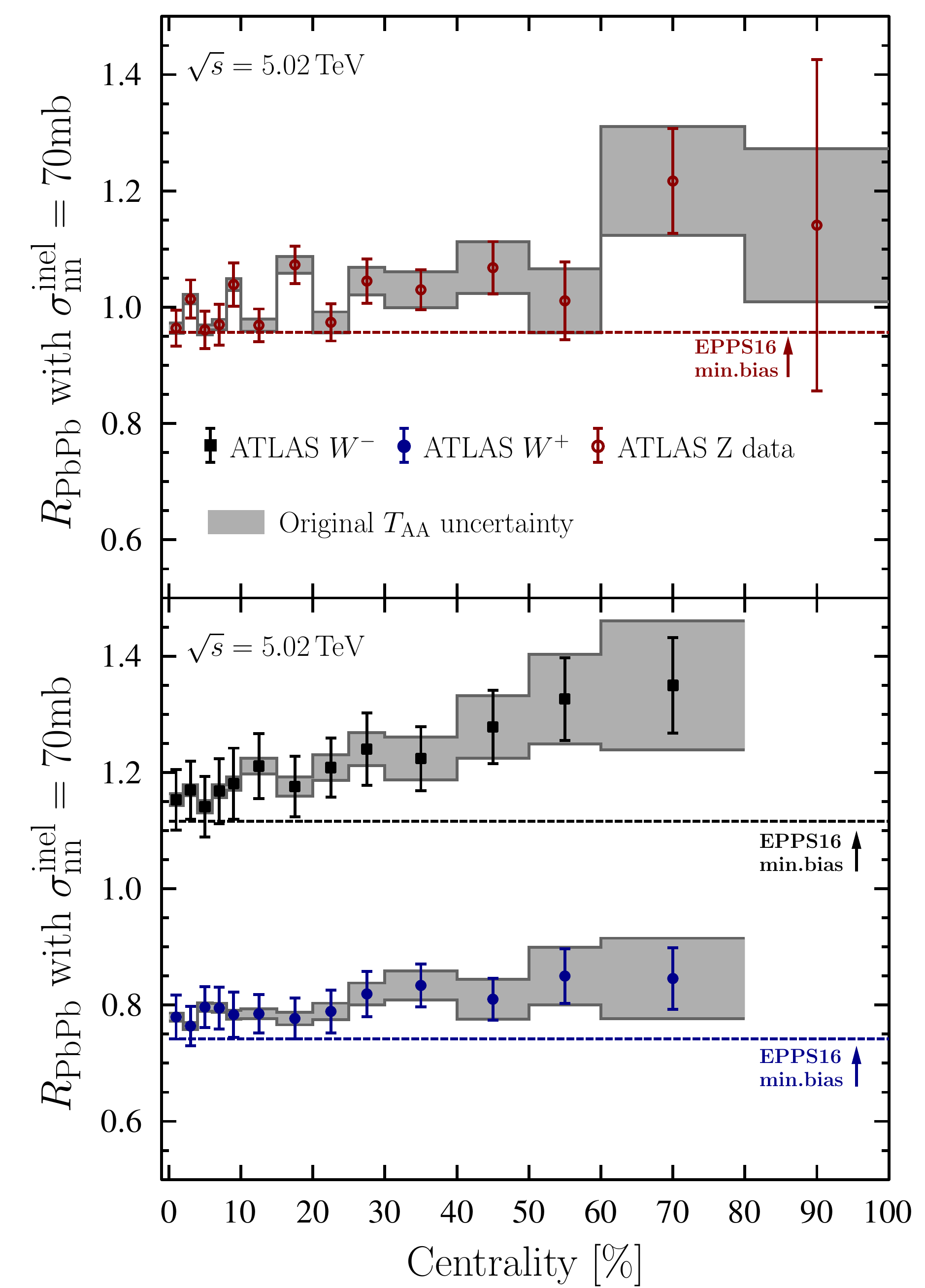}
\includegraphics[width=0.49\textwidth]{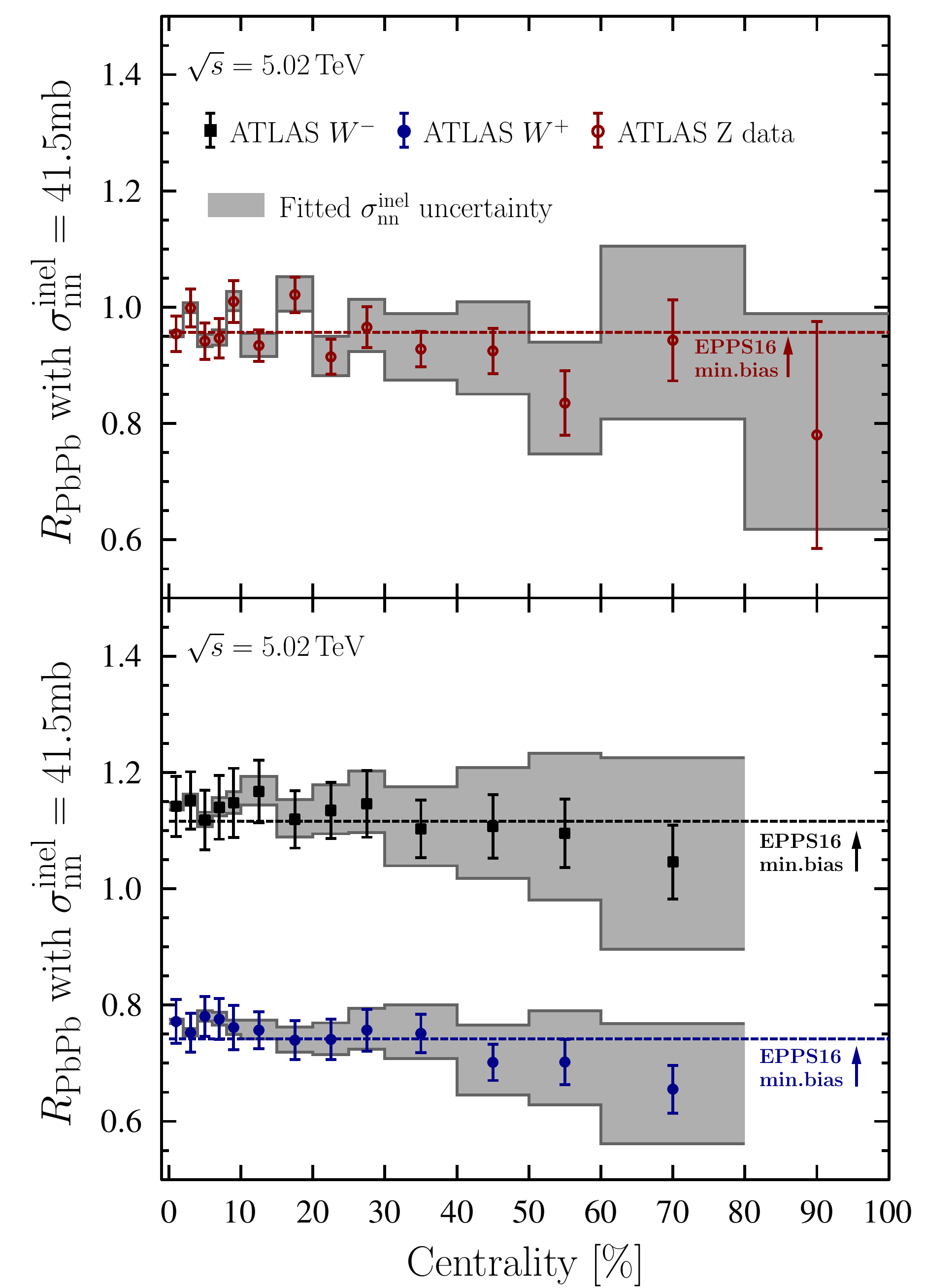}
\caption{The centrality-dependent nuclear modification ratios for $W^{\pm}$ and $Z$ boson production in Pb+Pb collisions from ATLAS \cite{Aad:2019sfe, Aad:2019lan} compared to NNLO pQCD calculation with EPPS16 nuclear modification with the nominal value of $\sigma_{\mathrm{nn}}^{\mathrm{inel}} = 70.0~\text{mb}$ (left) and with the nuclear-suppressed value $\sigma_{\mathrm{nn}}^{\mathrm{inel}} = 41.5~\text{mb}$ (right).}
\label{fig:R_PbPb_cent}
\end{figure*}


\section{Centrality dependence}

Even the quite significant suppression in $\sigma_{\mathrm{nn}}^{\mathrm{inel}}$ leads to rather modest modifications in $\langle T_{AA} \rangle$ for central and (close-to) minimum-bias collisions. The impact, however, grows towards more peripheral centrality classes, see Table~\ref{tab:cent}. To illustrate this, Fig.~\ref{fig:R_PbPb_cent} compares the centrality dependent $R_{\mathrm{PbPb}}^{\mathrm{exp}}$  before and after rescaling the data by ${\langle T_{AA}(\sigma_{\mathrm{pp}}^{\mathrm{inel}})\rangle}/{\langle T_{AA}(\sigma_{\mathrm{nn}}^{\mathrm{inel}})\rangle}$ using the fitted $\sigma_{\mathrm{nn}}^{\mathrm{inel}}$. The left-hand panels show the original ATLAS data including the quoted $\langle T_{AA} \rangle$ uncertainties, and in the right-hand panels the data have been rescaled and the uncertainties follow from the $\sigma_{\mathrm{nn}}^{\mathrm{inel}}$ fit. The striking effect is that the mysterious rise towards more peripheral collisions in the original data becomes compatible with a negligible centrality dependence, the central values indicating perhaps a mildly decreasing trend towards peripheral bins. As discussed e.g. in the ATLAS publications \cite{Aad:2019sfe, Aad:2019lan}, such a suppression could be expected from selection and geometrical biases associated with the MC Glauber modeling \cite{Morsch:2017brb}. Also other effects such as possible centrality dependence of $\sigma_{\mathrm{nn}}^{\mathrm{inel}}$ and the neutron-skin effect \cite{Paukkunen:2015bwa,Helenius:2016dsk} may become relevant to explain the data behaviour in the far periphery. 


\section{Minijets with shadowing}

\begin{figure}
\includegraphics[width=0.49\textwidth]{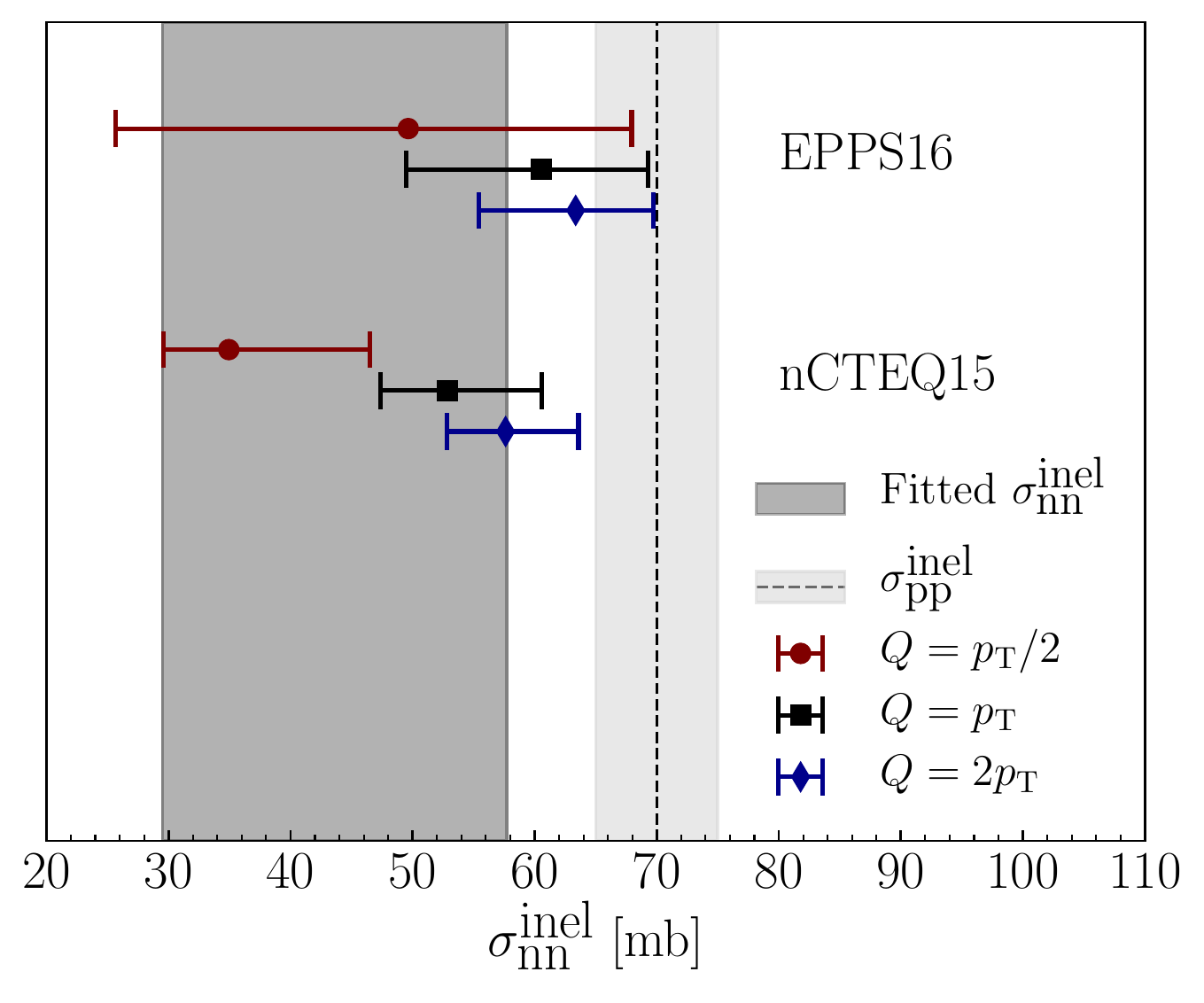}
\caption{
Predictions for $\sigma_{\mathrm{nn}}^{\mathrm{inel}}$ from an eikonal minijet model with the EPPS16 and nCTEQ15 nuclear PDFs for three scale choices. The fitted $\sigma_{\mathrm{nn}}^{\mathrm{inel}}$ is indicated with the dark-gray band and the nominal $\sigma_{\mathrm{pp}}^{\mathrm{inel}}$ with the light-gray band.
}
\label{fig:eikonal}
\end{figure}

To study the plausibility of the obtained suppression in $\sigma_{\mathrm{nn}}^{\mathrm{inel}}$, we calculate its value in an eikonal model for minijet production with nuclear shadowing. The model is based on a similar setup as in Ref.~\cite{Wang:1990qp} but in the eikonal function we include only the contribution from the hard minijet cross section $\sigma_{\mathrm{jet}}(\sqrt{s_{\mathrm{nn}}}, p_0, [Q])$, calculated at leading order in pQCD. The transverse-momentum cutoff $p_0$ (which depends on $\sqrt{s_{\mathrm{nn}}}$, scale choice $Q$ and the proton thickness) and the width of the assumed Gaussian proton thickness function we fix so that the model reproduces $\sigma_{\mathrm{pp}}^{\mathrm{inel}}=70$~mb matching the COMPETE analysis \cite{Cudell:2002xe} at $\sqrt{s} = 5.02~\text{GeV}$. The free proton PDFs are here \textsc{CT14lo} \cite{Dulat:2015mca}, and we take the nuclear PDF modifications from the EPPS16 \cite{Eskola:2016oht} and nCTEQ15 \cite{Kovarik:2015cma} analyses. The results for $\sigma_{\mathrm{nn}}^{\mathrm{inel}}$, obtained with $p_0$ and proton thickness function width fixed to the the p+p case, are shown in Fig.~\ref{fig:eikonal}. The error bars are again from the nuclear PDFs scaled to the 68\% confidence level. As expected at the few-GeV scales, the predicted $\sigma_{\mathrm{nn}}^{\mathrm{inel}}$ depends strongly on the factorization/renormalization scale $Q$, but within the uncertainties the nuclear suppression obtained from the fits to the ATLAS $W^{\pm}$ and $Z$ data seems compatible with the eikonal model predictions with both nuclear PDFs.


\section{Summary}

In the canonical approach the normalization for the measured per-event yields in nuclear collisions is obtained from the Glauber model taking the value of $\sigma_{\mathrm{nn}}^{\mathrm{inel}}$ from proton-proton measurements. Contrary to this, our strategy was to compare the state-of-the-art pQCD calculations with the measured $W^{\pm}$ and $Z$ boson $R_{\mathrm{PbPb}}$ and thereby unfold the value for $\sigma_{\mathrm{nn}}^{\mathrm{inel}}$ at $\sqrt{s_{\mathrm{nn}}} = 5.02~\text{TeV}$. We find that the recent high-precision ATLAS data from Run II prefer the value $\sigma_{\mathrm{nn}}^{\mathrm{inel}} = 41.5^{+16.2}_{-12.0}~\text{mb}\,,$ which is significantly lower than $\sigma_{\mathrm{pp}}^{\mathrm{inel}} = 70\pm 5~\text{mb}$. Such a suppression is in line with the expectations from an eikonal minijet model including nuclear shadowing. Remarkably, when using the fitted value for $\sigma_{\mathrm{nn}}^{\mathrm{inel}}$, the unexpected enhancements of $R_{\mathrm{PbPb}}$ in peripheral collisions disappear and the results become compatible with no centrality dependence. A possible hint of a slight decreasing trend toward peripheral collisions is observed which would be qualitatively in line with possible selection and geometrical biases. Our results thus suggest that the standard paradigm of using $\sigma_{\mathrm{pp}}^{\mathrm{inel}}$ as an input to Glauber modeling potentially leads to a misinterpretation of the experimental data.

\begin{acknowledgments}
We thank Mirta Dumancic for discussion concerning the ATLAS measurements. We acknowledge the Academy of Finland, projects 297058 (K.~J.~E.) and 308301 (H.~P. and I.~H.), and the V\"ais\"al\"a Foundation (M.~K.) for financial support. The Finnish IT Center for Science (CSC) is acknowledged for the computing time through the project jyy2580.
\end{acknowledgments}

\bibliographystyle{apsrev4-1}
\bibliography{GlauberPbPb_aps}

\end{document}